\documentclass[12pt]{article}

\def\PI     {{\rm P1}}
\def\PII    {{\rm P2}}
\def \D {\hbox{d}}
\def \Oeuvres{O$\!$euvres}

\begin{document}

\title{Introduction to the Painlev\'e property, test and analysis\footnote
 {Tampa, 9--11 March 2013 \hfill 5~September~2013}}

\author{Robert Conte\dag\ and M.~Musette\ddag\
{}\\
\\ \dag
LRC MESO,
\\ \'Ecole normale sup\'erieure de Cachan (CMLA) et CEA--DAM
\\ 61, avenue du Pr\'esident Wilson,
\\ F--94235 Cachan Cedex, France.
\\and
\\ Department of Mathematics, The University of Hong Kong
\\ Pokfulam, Hong Kong
\\ E-mail:  Robert.Conte@cea.fr
{}\\
\\ \ddag Dienst Theoretische Natuurkunde,
Vrije Universiteit Brussel,
Pleinlaan 2,
\\
\noindent
B-1050 Bruxelles
\\ E-mail: MMusette@vub.ac.be
}

\maketitle

\begin{abstract}
This short survey presents the essential features of what is called Painlev\'e analysis,
i.e.~the set of methods based on the singularities of differential equations
in order to perform their explicit integration.
Full details can be found in \textit{The Painlev\'e handbook}
or in various lecture notes posted on arXiv.
\end{abstract}



\section{Motivation. Definition of the Painlev\'e property}
\indent

Differentiating is easy, but the converse is not:
performing the integration of a given algebraic ordinary differential equation (ODE)
requires methods, not tricks. 
The main observation is: 
whenever one succeeds to find a closed form expression for the general solution $u(x)$ of an ODE,
this expression only involves solutions of other, usually lower order ODEs.
For instance, since the $\Gamma$ function has been proven 
\cite{Hoelder} to obey no differential equation, 
it can never contribute. Hence the

\textbf{Motivation} (L.~Fuchs, Poincar\'e): to define new functions by ODEs.

A function being a one-to-one application is by definition a singlevalued object.

Linear ODEs are all capable to define a function because their general solution
has no singularity which depends on the constants of integration. 

What about nonlinear ODEs? 
Their ability to define a function or not only relies on the nature of the singularities of 
their general solution.
There exist two classifications of singularities of the solutions of ODEs: 
on one hand to be critical or noncritical (i.e.~to display local multivaluedness or local singlevaluedness
around a singularity),
on the other hand to be fixed (their location does not depend on the initial conditions) or movable (the opposite).
See examples in Table \ref{TableSingularities}.

\tabcolsep=1.5truemm
\tabcolsep=0.5truemm

\begin{table}[h] 
\caption 
{
Examples of the four configurations of singularities.
The location $c$ depends on the constants of integration (i.e.~on the initial conditions) 
}
\vspace{0.2truecm}
\begin{tabular}{| c | c | c |}
\hline 
     & critical & noncritical 
\\ \hline  
movable   & $(x-c)^{1/2}$        & $e^{1/(x-c)}$        
\\ \hline  
fixed   & $\log(x-1)$ & $(x-2)^{-3}$ 
\\ \hline  
\end{tabular}
\label{TableSingularities}
\end{table}

For singlevaluedness, the only worry is that a singularity be at the same time movable and critical,
since one then does not know where to put a cut in order to remove the multivaluedness. Hence the

\textbf{Definition}. \textbf{Painlev\'e property} (PP) of an ODE: 
its general solution has no movable critical singularities.

Any other definition, such as ``All solutions have only movable poles'', is incorrect,
see Chazy (1911) and \cite[FAQ section]{CMBook}. 

Warning: essential singularities are \textit{not} involved in the definition of the PP.

The group of invariance of the PP
is the class of transformations 
(homographic on the dependent variable, analytic on the independent variable)
\begin{eqnarray}
& &
(u,x) \mapsto (U,X),\
u(x)=\frac{\alpha(x) U(X) + \beta(x)}{\gamma(x) U(X) + \delta(x)},\
X=\xi(x),\
\nonumber
\\
& &
 (\alpha, \beta, \gamma, \delta, \xi) \hbox{ functions},\
\alpha \delta - \beta \gamma \not=0,
\label{eqHomographicGroup}
\end{eqnarray}
in which $ \alpha, \beta, \gamma, \delta, \xi$ are arbitrary analytic 
functions. It depends on four arbitrary functions.

In order to find new functions, one must 
(i) investigate nonlinear ODEs of order one, then two, then three \dots;
(ii) select those which possess the PP,
(iii) prove whether their general solution defines a new (or an old) function.

This ambitious program yields various intermediate outputs,
which we now outline.

\section{By-products of the definition of the PP}
\indent

\begin{enumerate}
\item
\textbf{Painlev\'e test}. This term denotes a collection of algorithms,
detailed in section III, 
which provide necessary conditions for the PP.

Warning: these conditions are \textit{a priori} not sufficient, 
i.e.~passing the Painlev\'e test does \textit{not} imply possessing the PP.
Indeed, as an algorithm, the test must terminate after a finite number of steps.
In 1893, Picard built the following example to illustrate this point:
denoting $\wp$ the elliptic function of Weierstrass defined 
by Eq.~(\ref{eqdefwp}) below,
the variable $u(x)=\wp(\lambda \log(x-c_1)+c_2,g_2,g_3)$
obeys (by the elimination of $c_1,c_2$) an algebraic second order ODE and
is singlevalued iff $2 \pi i \lambda$ is a period of $u$,
a transcendental condition on $\lambda,g_2,g_3$ impossible to obtain in a finite number
of algebraic steps.

\item
\textbf{Painlev\'e analysis}. By extension, this denotes all the methods based on singularities,
whose aim is to generate any kind of closed-form result (particular solution,
first integral, Darboux polynomial, Lax pair, etc). More in \cite{CM2003}. 

\item
\textbf{Classifications}. 
This denotes the exhaustive lists of ODEs in a certain class (e.g.~third order second degree)
which have the PP and are explicitly integrated.
Established after a lot of work (mainly by Painlev\'e, Gambier, Chazy, Bureau, Exton, Martynov, Cosgrove,
see the state of the art in \cite[Appendix A]{CMBook}),
they serve as tables to help integrating a given nonlinear ODE,
and the method to proceed is described, for instance, in \cite[\S 3.1.1.2]{CMBook}. 

Let us just give one example, taken from the Lorenz model \cite{Segur},
detailed in \cite[\S 3.1.1.2]{CMBook}.
The second order ODE for $x(t)$
\begin{eqnarray}
& &
\frac{\D^2 x}{\D t^2}
+ 2 \frac{\D x}{\D t}
+ \frac{x^3}{2} + \left( \frac{8}{9} - \frac{K_1}{2} e^{-2 t} \right) x=0,
\label{eqLorenz2119}
\end{eqnarray}
passes the Painlev\'e test and
cannot be made autonomous by a transformation
(\ref{eqHomographicGroup}).
In order to integrate,
the strategy is to map it to one nonautonomous equation
among the fifty equations
of the list of Gambier \cite{GambierThese,Ince,Davis}.
In order to decide which equation is suitable in this list,
the method \cite{PaiBSMF} is to write (\ref{eqLorenz2119})
in the canonical form
\begin{eqnarray}
& &
\frac{\D^2 x}{\D t^2} =
A_2(x,t) \left(\frac{\D x}{\D t}\right)^2 + A_1(x,t) \frac{\D x}{\D t}
+A_0(x,t).
\label{eqA2A1A0}
\end{eqnarray}
If the ODE has the Painlev\'e property,
as shown by Painlev\'e \cite[p.~258]{PaiBSMF} \cite[p.~74]{PaiActa},
the coefficient $A_2$ must be the sum of at most four simple poles in $x$
(possibly including $x=\infty$),
whose sum of residues is equal to two
\begin{eqnarray}
& &
A_2(x,t) = \sum_j \frac{r_j}{x-a_j(t)},\ \sum_j r_j=2,
\end{eqnarray}
and the number of poles and the set of residues $\lbrace r_j\rbrace$
is invariant under the homographic transformations (\ref{eqHomographicGroup}).
In the present case,
the value $A_2=0$
means the single pole $x=\infty$ with residue $2$
(in order to check it, change $x \to 1/x$).
The method is
then to select the few Gambier equations 
whose coefficient $A_2$ has the same
number of poles and the same set of residues $\lbrace r_j\rbrace$,
here one pole with a residue two,
finally to map the ODE (\ref{eqLorenz2119})
to one of these Gambier equations by a transformation
(\ref{eqHomographicGroup}).
Here the suitable Gambier equation is
either $\PII$ or $\PI$.
To choose between $\PII$ and $\PI$,
one also matches the structure of movable singularities
(two simple poles with opposite residues for $\PII$ and $x(t)$,
one double pole for $\PI$),
which now selects $\PII$ as the unique possible match.
Indeed, the particular transformation (\ref{eqHomographicGroup})
\begin{eqnarray}
& &
x=a e^{-2 t/3} X,\ T=\frac{i}{2} a^{-3/2}e^{-2 t/3},\
K_1=\frac{3}{8} i a^3,
\end{eqnarray}
maps (\ref{eqLorenz2119}) to
\begin{eqnarray}
& &
\frac{\D^2 X}{\D T^2} = 2 X^3 + T X + \alpha,\ \alpha=0.
\end{eqnarray}
\index{$\PII$!equation}

\item
\textbf{New functions}.
The main difficulty is to prove the irreducibility (to a linear ODE
or to a nonlinear, lower order ODE).

Order one. First order algebraic ODEs define one and only one new function, the elliptic function.
We say ``the'' because all elliptic functions are deduced from some canonical representative
such as the Weierstrass function $\wp$, see (\ref{eqbiraelliptic}) and (\ref{eqdefwp}).

Order two. This defines only six functions, the Painlev\'e functions (Painlev\'e 1900, R.~Fuchs 1905).

Order three. No new function.

Order four. There exist at least five first degree ODEs 
(labeled F-V, F-VI, F-XVII, F-XVIII, Fif-IV in Cosgrove \cite{CosPole2,CosPole1})
which have a singlevalued general solution 
with a transcendental dependence on the four constants of integration.
However, their irreducibility is unsettled yet.
To have an idea of the difficulty of the proof,
consider the ODE, easy to establish by elimination,
whose general solution is $u(x)=u_1(x)+u_2(x)$,
where $u_j(x)$ obeys the first Painlev\'e equation.
It has by construction the PP,
a transcendental dependence on the four constants of integration,
but a reducible general solution.
One must therefore be very cautious about this question of irreducibility.

\end{enumerate}

\section{The Painlev\'e test}
\label{sectionPtest}
\indent

The main message is that the Painlev\'e test does \textit{not} reduce to 
the method of Kowalevski (1889) and Gambier (1910),
popularized by Ablowitz, Ramani and Segur \cite{ARS1978}.

The main methods of the test are

\begin{enumerate}

\item
The \textbf{$\alpha$-method of Painlev\'e} (1900).
This is the most powerful, but it involves solving differential equations,
while all other methods involve solving algebraic equations. 
Tutorial presentation in \cite[\S 5.5]{Cargese1996Conte}

\item
The \textbf{method of Kowalevski and Gambier},
later made rigorous by Bureau (1939, 1964).
This well known method consists in requiring the existence,
near every movable singularity,
of a Laurent series able to represent the general solution.
We will not detail it here because it can be found in many courses,
such as \cite{Cargese1996Conte},
but we will rather concentrate on common errors and difficulties.

Two errors should not be made:

\textit{Error 1}. 
Discard negative integer Fuchs indices 
(also called resonances, or Kowalevski exponents,
these are all synonyms)
as containing no information.
Example: the ODE
\begin{equation}
 u'''' + 3 u u'' - 4 u'^2 = 0,
\label{eqBureauOrder4}
\end{equation}
admits near a movable singularity $x_0$ the two families
\begin{eqnarray}
& &
{\hskip -16.0 truemm}
u \sim -60 /(x-x_0)^2,\ \hbox{Fuchs indices } -3,-2,-1,20,
\\
& &
{\hskip -16.0 truemm}
u \sim u_0 /(x-x_0)^3,\ u_0 \hbox{ arbitrary, Fuchs indices } -1,0,
\label{eqBureau4p3}
\end{eqnarray}
among which the first family has negative indices $-3,-2$.
This ODE is processed below.

\textit{Error 2}. 
Interpret a negative integer Fuchs index as the presence of an essential singularity.
Example: the ODE
\begin{equation}
 u'' + 3 u u' + u^3 = 0
\end{equation}
admits the family $u \sim 2 (x-x_0)^{-1}$
with the Fuchs indices $-2,-1$,
and it has no essential sigularity because its general solution
$u=\frac{1}{x-a} + \frac{1}{x-b}$,
in which $a,b$ are the two constants of integration, has only poles.

There exist two situations making the method of Kowalevski and Gambier
indecisive:

\begin{enumerate}
\item
presence of negative integers among the set
of Fuchs indices,

\item
insufficient number of Fuchs indices,
i.e.~lower than the differential order of the ODE.

\end{enumerate}


\item
\textbf{Fuchsian perturbative method} \cite{FP1991,CFP1993}.
It deals with the first indecisive situation (negative integer Fuchs indices).

It considers the Laurent series of Gambier as the zero-th order of a Taylor series in a small
parameter $\varepsilon$ and requires singlevaluedness at each order in $\varepsilon$.
With the first family of (\ref{eqBureau4p3}),
a movable logarithm arises from the Fuchs index $-1$ at seventh order of perturbation.


\item
\textbf{NonFuchsian perturbative method} \cite{MC1995}.
This handles the second indecisive situation (number of Fuchs indices lower than the differential order),
provided one knows a particular solution in closed form.

The Laurent series of the second family of (\ref{eqBureau4p3}) happens to terminate and defines the
closed form two-parameter particular solution
\begin{equation}
 u^{(0)}=c (x-x_0)^{-3}-60(x-x_0)^{-2},\ (c,x_0) \hbox{ arbitrary}.
\label{eqBureau4PartSol}
\end{equation}
Under the perturbation
$u=u^{(0)} + \varepsilon u^{(1)} + \dots$,
the ODE for $u^{(1)}$ 
\begin{equation}
E^{(1)} = E'(x,u^{(0)}) u^{(1)} \equiv
 [              \partial_x^4
  + 3 u^{(0)}   \partial_x^2
  - 8 u^{(0)}_x \partial_x
  + 3 u^{(0)}_{xx}] u^{(1)} = 0,
\label{eqBureauLin}
\end{equation}
is known globally (not only locally near $x=x_0$) because $u^{(0)}$ is closed form,
therefore one can test all its singularities for singlevaluedness.
Testing $x=x_0$ (nonFuchsian) is what the Fuchsian perturbative method has done above.
Testing $x=\infty$ (Fuchsian) immediately uncovers a movable logarithm \cite{MC1995}
arising from a Neumann function.

\end{enumerate}

\section{Methods for finding all elliptic solutions of a given ODE}
\indent

Quite a number of presentations in this Tampa conference 
deal with finding, by various methods, some particular solutions of a given $N$-th order nonlinear ODE
in the class of either elliptic functions or rational functions of one exponential
or rational functions.
\medskip

What we would like to point out is the existence of two methods, turned into algorithms,
to find not only \textit{some} but \textit{all} the particular solutions in that class
(elliptic and degenerate of elliptic).
Not surprisingly, both methods take advantage of the Laurent series. 
Accordingly, these two algorithms \textit{must} be used,
under penalty of missing some solutions.
\medskip

Let us first recall some definitions.

\textbf{Definition}.
A function $u$ of a complex variable $x$ is called \textit{elliptic} iff it is meromorphic and doubly periodic.

Given two arbitrary elliptic functions $u,v$, they are birationally equivalent,
i.e.~there exist two rational functions $R_1,R_2$ such that
\begin{eqnarray}
& &
u=R_1(v,v'),\ v=R_2(u,u').
\label{eqbiraelliptic}
\end{eqnarray}
Let us denote $\wp$ the canonical elliptic function introduced by Weierstrass,
defined by the first order ODE
\begin{eqnarray}
& &
{\wp'}^2= 4 \wp^3 - g_2 \wp - g_3
        =4(\wp(x)-e_1)(\wp(x)-e_2)(\wp(x)-e_3).
\label{eqdefwp}
\end{eqnarray}

Choosing $v=\wp$ in (\ref{eqbiraelliptic}),
one obtains the two successive degeneracies of elliptic functions,
\begin{itemize}
\item
when one root $e_j$ is double ($g_2^3-27 g_3^2=0$),
degeneracy to simply periodic meromorphic functions
(i.e.~rational functions of one exponential $e^{k x}$,
which includes usual trigonometric and hyperbolic trigonometric functions)
according to
\begin{eqnarray}
& &
\forall x,d:\
\wp(x,3 d^2,-d^3)
 = - d         + \frac{3 d}{2} \coth^  2  \sqrt{\frac{3 d}{2}} x,
\label{eqwpcoth}
\end{eqnarray}

\item
when the root $e_j$ is triple ($g_2=g_3=0$),
degeneracy to rational functions of $x$.

\end{itemize}

\subsection{First method}

The first method \cite{MC2003,CMBook,CM2009} 
implements a classical theorem of Briot and Bouquet,
stating that any such solution obeys a first order algebraic ODE
$F(u',u)=0$ in which the degrees of the polynomial $F$ in $u$ and $u'$ are known
and obtained from the given $N$-th order ODE by carefully counting its number of movable poles.
\medskip

Example.

Consider the ODE
\begin{equation}
u''=6 u^2 -\frac{a}{2}.
\label{eqcobaye1}
\end{equation}
One first counts carefully all its movable poles.
 The result here is that $u$ has one movable double pole
\begin{equation}
u=\frac{1}{\chi^2} + \dots,\ \chi=x-x_0.
\label{eqLaurent1}
\end{equation}  
Assuming that some solution $u$ of (\ref{eqcobaye1}) is nondegenerate elliptic,
one then computes the \textit{elliptic order} of $u$ and $u'$,
i.e.~their total number of poles, counting multiplicity.
The result is $m=2$ for $u$ and $n=3$ for $u'$.
The above mentioned theorem of Briot and Bouquet states that,
if $u$ is nondegenerate elliptic,
there exists a polynomial $F$ of $(u,u')$
of degree $m$ in $u'$ and $n$ in $u$,
such that
\begin{equation}
F(u,u')=0.
\label{eqcobaye1P}
\end{equation}
In our example, this polynomial (which must also obey other constraints)
has the necessary form
\begin{equation}
P={u'}^2 +(b_1 u + b_0) u'+ a_3 u^3 + a_2 u^2 + a_1 u + a_0.
\label{eqpolP}
\end{equation}
One then requires that the Laurent series (\ref{eqLaurent1}) also obeys (\ref{eqpolP}).
This generates a linear system in the unknowns $a_k,b_k$,
quite easy to solve because it is overdetermined.
One has thus reduced the search of ellliptic solutions of a $N$-th order ODE to the integration
of a first order ODE.
This latter problem can be solved for instance by the computer algebra package \textit{algcurves} of Maple.

\subsection{Second method}

The second method \cite{DK2011method,DK_ODE3} 
implements another classical result of Hermite \cite{Hermite1888},
stating that any elliptic or degenerate elliptic function admits a unique decomposition in simple elements
\begin{equation}
u(x)=C+P(x)+\sum_{k=1}^M \sum_{j=0}^{m_k} c_{k,j}\ \zeta^{(j)}(x-a_k),\ 
\label{eqHermite}
\end{equation}
in which 
the integers $M$ and $m_k$ are again provided by counting the movable poles,
$C,c_{k,j},a_k$ are complex constants,
$\zeta$ is the function of Weierstrass ($\zeta'=-\wp$)
and $P$ is a polynomial 
which is nonzero only for degenerate elliptic solutions
(its degree is one for simply periodic solutions,
and equal to the order of the pole $x=\infty$ for rational solutions).
\medskip

Like in the first method, one computes all Laurent series up to some finite order.
The singular parts of these $M$ series determine completely the coefficients $c_{k,j}$ in (\ref{eqHermite}).
In order to then determine $C$ and $a_k$, one uses the addition formula of the elliptic functions
to write (\ref{eqHermite}) as a rational function of $\wp(x-a_1),\wp'(x-a_1)$, see Eq.~(\ref{eqbiraelliptic}).
Inserting in this last expression the Laurent series of $u$ near $x=a_1$, one generates 
algebraic relations involving $\wp(a_k-a_1), k>1$, whose solution yields the desired result.

The previous example (\ref{eqcobaye1}) (one double pole) leads to 
\begin{equation}
u(x)=C+\zeta^{(1)}(x-a_1),\ 
\label{eqHermite1}
\end{equation}
and the computation will provide $C=0, g_2=a,g_3=\hbox{arbitrary}$.

The example
\begin{equation}
{u'}^2=(u-a)(u-b)(u-c)(u-d)
\label{eqcobaye2}
\end{equation}
yields the necessary form of the elliptic solution
\begin{equation}
u(x)=C+\zeta(x-a_1)-\zeta(x-a_2),\ 
\label{eqHermite2}
\end{equation}
and the remaining unknowns $\wp(a_2-a_1),\wp'(a_2-a_1),g_2,g_3$
are obtained by expanding (\ref{eqHermite2}) in Laurent series near $x=a_1$
and identifying this expansion with the Laurent series near $x=a_1$ computed from (\ref{eqcobaye2}).

Because they can find all elliptic and degenerate elliptic solutions, 
these two methods make obsolete all the previous ones.



\textbf{Acknowledgments}
  One of us (RC) is quite happy to acknowledge the generous support of the organizers.

\end{document}